\documentclass[english,pra,twocolumn]{revtex4-1}
\usepackage[T1]{fontenc}
\usepackage[latin9]{inputenc}
\usepackage{color}
\usepackage{amsmath}
\usepackage{amssymb}
\usepackage{stackrel}
\usepackage{graphicx}

\begin{document}

\title{Nonreciprocal light transmission via optomechanical parametric interactions}

\author{Yan-Ting Lan}
\affiliation{Fujian Key Laboratory of Quantum Information and Quantum Optics and
Department of Physics, Fuzhou University, Fuzhou 350116, People's
Republic of China}

\author{Wan-Jun Su}
\affiliation{Fujian Key Laboratory of Quantum Information and Quantum Optics and
Department of Physics, Fuzhou University, Fuzhou 350116, People's
Republic of China}

%\author[1]{Huaizhi Wu}
\author{Huaizhi Wu}
\affiliation{Fujian Key Laboratory of Quantum Information and Quantum Optics and
Department of Physics, Fuzhou University, Fuzhou 350116, People's
Republic of China}

\author{Yong Li}
\affiliation{Beijing Computational Science Research Center, Beijing 100193, People's
Republic of China}

%\author[1,*]{Shi-Biao Zheng}
\author{Shi-Biao Zheng}
\affiliation{Fujian Key Laboratory of Quantum Information and Quantum Optics and
Department of Physics, Fuzhou University, Fuzhou 350116, People's
Republic of China}

%\affil[*]{Corresponding author: huaizhi.wu@fzu.edu.cn}
\begin{abstract}

Nonreciprocal transmission of optical or microwave signals is indispensable
in various applications involving sensitive measurements. In this
paper, we study optomechanically induced directional amplification
and isolation in a generic setup including two cavities and two mechanical
oscillators by exclusively using blue-sideband drive tones. The input
and output ports defined by the two cavity modes are coupled through
coherent and dissipative paths mediated by the two mechanical resonators,
respectively. By choosing appropriate transfer phases and strengths
of the driving fields, either a directional amplifier or an isolator
can be implemented at low thermal temperature, and both of them show
bi-directional nonreciprocity working at two mirrored frequencies.
The nonreciprocal device can potentially be demonstrated by opto-
and electro-mechanical setups in both optical and microwave domains. 
\end{abstract}

\maketitle

Nonreciprocal devices, such as isolators and directional amplifiers,
are essential in communication and signal processing, as they protect
the signal source from disturbances by the measurements and extraneous
noises. A typical way to achieve nonreciprocal transmission of the
signal is to divide the signal into two branches and interfere differently
for the forward and the backward propagation directions by a controlled
phase difference \cite{Koch2010,Kamal2011}. Conventional realization
of optical and microwave nonreciprocity is built on addressing magnetic-field
induced effects, which are hard to integrate on chips \cite{Haldane2008,Khanikaev2010,Bi2011}.
A more suitable implementation of integrated devices can be realized
with superconducting quantum circuits, where the strong Josephson
nonlinearity and parametric pumping have been exploited to realize
circulators, directional amplifiers \cite{Kamal2011,Abdo2014,Sliwa2015,Lecocq2017}
and isolators \cite{Abdo2019}, but working at microwave frequencies. 

Recently, optomechanics \cite{Aspelmeyer2014} becomes a new promising
platform for realizing nonreciprocity via multi-path interference
\cite{Xu2015,Ranzani2015,Shen2016,Tian2017,Peterson2017,Malz2018,Huang2021}.
In recent theoretical work, Metelmann and Clerk have proposed that
any coherent interaction can be made directionally by reservoir engineering
an auxiliary dissipative path \cite{Metelmann2015}. The idea has
found good implementations in both asymmetric three-mode and symmetric
four-mode optomechanical setups, basing on which nonreciprocity in
the optical or microwave domain has been demonstrated with two mechanical
resonators each mediating both coherent and dissipative couplings
\cite{Bernier2017,Barzanjeh2017,Malz2018}. In general, optomechanically
induced directional amplification can be realized by including blue-sideband
drive tones \cite{Nunnenkamp2014,FangJ.Luo2017,Malz2018,Shen2018,Damskagg2019,MercierDeLepinay2020,Jiang2020}, and isolation \cite{Manipatruni2009,Xu2016,Ruesink2016,Bernier2017}
can be performed by using drive tones close to the red sidebands.
Moreover, optomechanical setups can be used as a nonreciprocal transducer,
interfacing microwave and optical elements \cite{Xu2016,Tian2017}.

Inspired by the previous schemes, in this paper, we focus on optomechanical
interactions inducing nonreciprocal transmission between two cavity
modes where two mechanical oscillators (MOs) respectively mediate
coherent and dissipative couplings on two transfer paths. Exclusive
use of blue-sideband drive tone leads to a linearized Hamiltonian,
involving only two-mode squeezing interactions. We find that the optomechanical
setup can be configured as both a directional amplifier and an isolator,
depending on the driving strengths and the global phase. In contrast
to Ref. \cite{FangJ.Luo2017}, where two optomechanical cavities are
driven by blue-detuned lasers with coherent optical (mechanical) couplings,
giving rise to a port-to-port net gain of 1 dB, the input and output
ports in our scheme are not directly connected and the optomechanically
induced net amplification can be larger than 11 dB. Remarkably, the
working points for both the directional amplifier and the isolator
do not appear at the cavity resonance. Indeed, due to the broken time-reversal
symmetry, the system shows a bi-directional nonreciprocal response
between two cavity modes at two different frequencies detuned from
the probe resonance by sub-linewidth. The scheme can potentially be
implemented in both the microwave and optical domains, and be used
for a microwave-to-optical transducer with the electromechanical systems.

\begin{figure}[htbp]
\includegraphics[width=0.95\columnwidth]{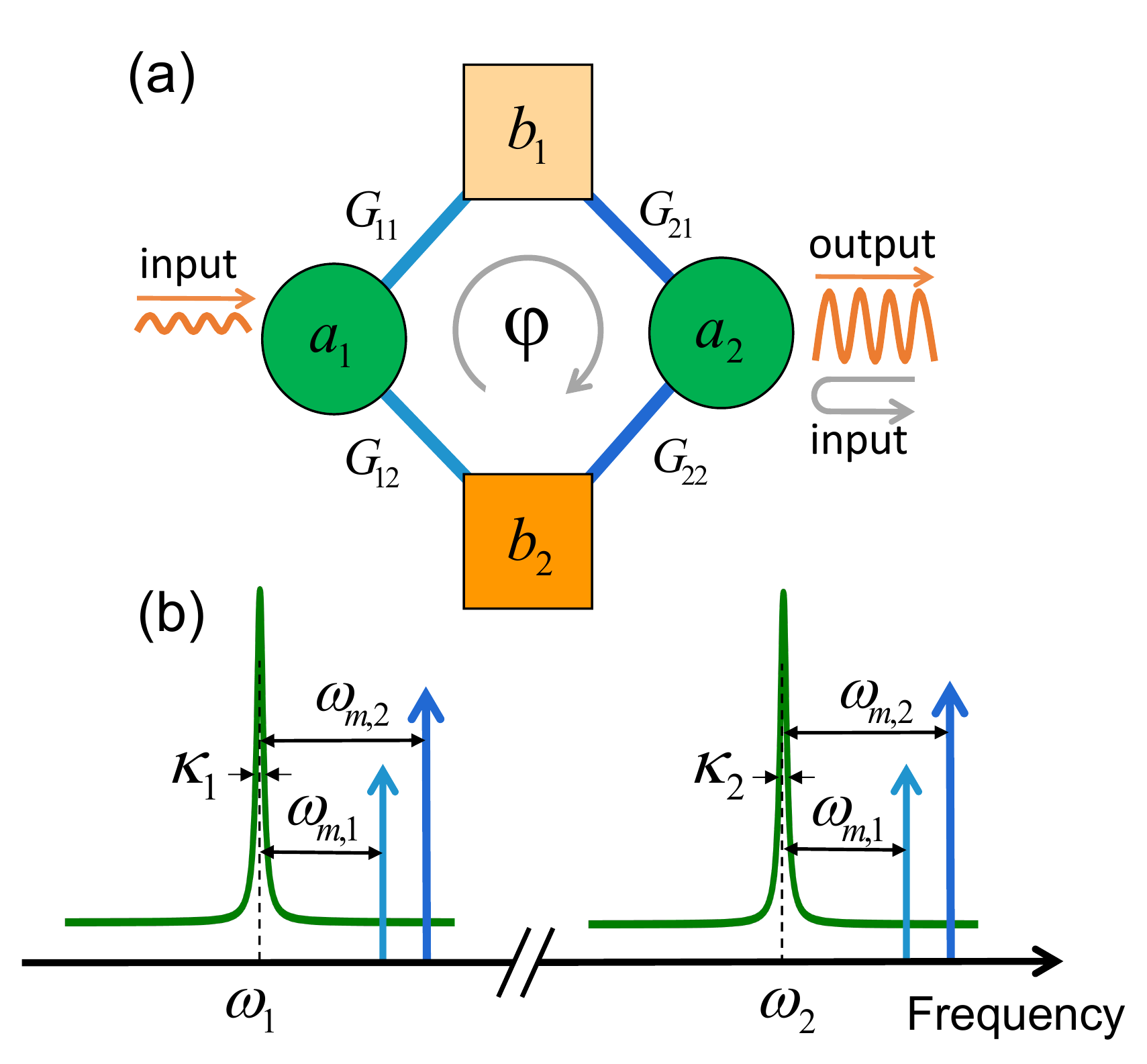}\caption{\label{fig:model1}Implementation of nonreciprocal transmission. (a)
Two cavity modes (denoted by $a_{1}$, $a_{2}$) independently couple
to two mechanical oscillators (denoted by $b_{1}$, $b_{2}$) via
optomechanical parametric interactions with the effective pairwise
coupling strengths $G_{i,j}$ ($i,j=1,2$).  $\varphi$ is the global phase 
introduced by the phase-correlated driving lasers. (b) Four-tone driving
scheme. The optomechanical cavities are driven at frequencies exactly
on the blue motional sidebands of the mechanical modes. }
\end{figure}
%\setboolean{displaycopyright}{true}
The schematic diagram of the device is shown in Fig. \ref{fig:model1}(a).
The setup consists of two cavity modes (with frequencies $\omega_{1}$
and $\omega_{2}$, and ring down rates $\kappa_{1}$ and $\kappa_{2}$,
respectively) acting as input and output ports, each of which is coupled
to two non-degenerate mechanical resonators (with motional frequencies
$\omega_{m,1}$ and $\omega_{m,2}$, and intrinsic damping rates $\gamma_{1}$
and $\gamma_{2}$). The Hamiltonian of the optomechanical system reads
($\hbar=1$) 

\begin{eqnarray}
H & = & \sum_{i=1,2}\omega_{i}a_{i}^{\dagger}a_{i}+\sum_{j=1,2}\omega_{m,j}b_{j}^{\dagger}b_{j}\nonumber \\
&  &+\sum_{i,j}g_{i,j}a_{i}^{\dagger}a_{i}(b_{j}^{\dagger}+b_{j}),\label{eq:system Hamiltonian}
\end{eqnarray}
where ($a_{i}^{\dagger}$,$a_{i}$)$_{i=1,2}$ are the cavity field
creation and annihilation operators, and ($b_{j}^{\dagger}$,$b_{j}$)$_{j=1,2}$
are the phononic operators. $g_{i,j}$ are the single-photon optomechanical
coupling strength between the cavity mode $i$ and the mechanical
mode $j$.\textcolor{black}{{} The model can be realized in microwave-
or electro-optomechanical systems, where the cavity modes for the
former are two microwave fields, while for the latter are a microwave
mode and an optical mode.}

The optomechanical cavities are respectively driven by two laser beams
on the well-resolved blue sidebands ($\omega_{m,j}\gg\kappa_{i},\gamma_{j}$)
with the frequencies $\omega_{i}+\omega_{m,j}$, see Fig. \ref{fig:model1}(b),
which induce two-mode squeezing interactions between the optical modes
and the mechanical modes \cite{Malz2018}, in contrast to previously
demonstrated optomechanical isolators that used exclusively drive
tones close to the red sidebands \cite{Xu2016,Peterson2017,Bernier2017}.
For strong drivings, the cavity field $a_{i}(t)$ ($i=1,2$) can be
decomposed into a coherent amplitude $\alpha_{i}(t)$ oscillating
at pump frequencies and a fluctuation part $\delta a_{i}(t)$. We
assume that min$[\omega_{m,j},|\omega_{m,1}-\omega_{m,2}|]\gg$max$[|g_{i,j}\alpha_{i}(t)|]$,
then the coherent parts can be approximately given by $\alpha_{i}(t)\thickapprox\sum_{j=1,2}\alpha_{i,j}e{}^{-i\omega_{m,j}t}$,
where $\alpha_{i,j}=|\alpha_{i,j}|e^{i\theta_{i,j}}$ are complex
number with $\theta_{i,j}$ relying upon the phase of the corresponding
laser pump. Using the standard semiclassical approach, we can obtain
the linearized Hamiltonian with the fluctuation operators $\delta a_{i}(t)$
($\delta b_{j}(t)$) {[}renamed as $a_{i}(t)$ ($b_{j}(t)$) for concision{]},
which in the frame rotating with $H_{0}=\sum_{i=1,2}\omega_{i}a_{i}^{\dagger}a_{i}+\sum_{j=1,2}\omega_{m,j}b_{j}^{\dagger}b_{j}$
reads

\begin{eqnarray}
H_{\text{Lin}} & = & G_{11}a_{1}b_{1}+G_{21}a_{2}^{\dagger}b_{1}^{\dagger}e^{i\varphi}+G_{22}a_{2}b_{2}\nonumber \\
 &  & +G_{12}a_{1}^{\dagger}b_{2}^{\dagger}+\text{H.c.},\label{eq:rotatingH}
\end{eqnarray}
where $G_{ij}=|g_{i,j}\alpha_{i,j}$| are the field-enhanced coupling
strengths and the terms oscillating at frequencies close to $\pm\left|\omega_{m,1}-\omega_{m,2}\right|$
and $2\omega_{m,j}$ are neglected (i.e. the rotating wave approximation).
Moreover, we have made gauge transformations to the operators $a_{1}\rightarrow a_{1}e^{i\theta_{12}}$,
$a_{2}\rightarrow a_{2}e^{i\theta_{22}}$ and $b_{1}\rightarrow b_{1}e^{i(\theta_{11}-\theta_{12})}$,
leaving only the global phase $\varphi=\theta_{21}-\theta_{22}-(\theta_{11}-\theta_{12})$
here, which can be addressed by using phase-correlated laser lights. 

The quantum Langevin equations (QLEs) for the fluctuation operators
are $\dot{\mu}=-M\mu+\sqrt{\Gamma}\mu_{\text{in}}$, where $\mu=(a_{1}^{\dagger},a_{2}^{\dagger},b_{1},b_{2})^{T}$,
$\Gamma=diag(\kappa_{1},\kappa_{2},\gamma_{1},\gamma_{2})$, 

\begin{eqnarray}
M & = & \left[\begin{array}{cccc}
\frac{\kappa_{1}}{2} & 0 & -iG_{11} & -iG_{12}\\
0 & \frac{\kappa_{2}}{2} & -iG_{21}e^{-i\varphi} & -iG_{22}\\
iG_{11} & iG_{21}e^{i\varphi} & \frac{\gamma_{1}}{2} & 0\\
iG_{12} & iG_{22} & 0 & \frac{\gamma_{2}}{2}
\end{array}\right],\label{eq:m matrix}
\end{eqnarray}
and $\mu_{\text{in}}=(a_{1,\text{in}}^{\dagger},a_{2,\text{in}}^{\dagger},b_{1,\text{in}},b_{2,\text{in}})^{T}$,
with $a_{i,\text{in}}^{\dagger}$ and $b_{j,\text{in}}$ being the
zero-mean noise operators for the cavities and mechanical oscillators,
satisfying the correlation functions $\langle a_{i,\text{in}}^{\dagger}(t)a_{i,\text{in}}(t^{\prime})\rangle=0$,
$\langle a_{i,\text{in}}(t)a_{i,\text{in}}^{\dagger}(t^{\prime})\rangle=\delta(t-t^{\prime})$,
$\langle b_{j,\text{in}}^{\dagger}(t)b_{j,\text{in}}(t^{\prime})\rangle=n_{m,j}\delta(t-t^{\prime})$,
$\langle b_{j,\text{in}}(t)b_{j,\text{in}}^{\dagger}(t^{\prime})\rangle=(n_{m,j}+1)\delta(t-t^{\prime})$
with the thermal phonon number $n_{m,j}=[\exp(\hbar\omega_{m,j}/k_{B}T)-1]^{-1}$
and negligible photonic thermal occupation. We then rewrite the QLEs
in the frequency domain with the Fourier transformations $o^{\dagger}(\omega)=\frac{1}{\sqrt{2\pi}}\int_{-\infty}^{+\infty}o^{\dagger}(t)e^{i\omega t}dt,$
leading to $\mu(\omega)=\left(M-i\omega I\right)^{-1}\sqrt{\Gamma}\mu_{\text{in}}(\omega),$
where $I$ is the identity matrix. Moreover, by combining the standard
input-output relation $\mu_{\text{out}}(\omega)=\mu_{\text{in}}(\omega)-\sqrt{\Gamma}\mu(\omega)$
\cite{Aspelmeyer2014,Gardiner1985} {[}with $\mu_{\text{out}}(\omega)=(a_{1,\text{out}}^{\dagger},a_{2,\text{out}}^{\dagger},b_{1,\text{out}},b_{2,\text{out}})${]}
and the noise spectrum approach $s_{o}(\omega)=\int_{-\infty}^{+\infty}d\omega^{\prime}\langle o^{\dagger}(\omega^{\prime})o(\omega)\rangle$,
we can obtain the spectrum of the output fields $S_{\text{out}}(\omega)=[s_{a_{1,\text{out}}}(\omega),s_{a_{2,\text{out}}}(\omega),s_{b_{1,\text{out}}}(\omega),s_{b_{2,\text{out}}}(\omega)]^{T}$,
which connect to the spectrum of input fields $S_{\text{in}}(\omega)=[s_{a_{1,\text{in}}}(\omega),s_{a_{2,\text{in}}}(\omega),s_{b_{1,\text{in}}}(\omega),s_{b_{2,\text{in}}}(\omega)]^{T}$
by the transmission matrix $T\left(\omega\right):$ 
\begin{equation}
S_{\text{out}}(\omega)=T(\omega)S_{\text{in}}(\omega),\label{eq:S_fullmodel}
\end{equation}
where the matrix elements $T_{ij}(\omega)=|S_{ij}(\omega)|^{2}$ (with
$S(\omega)=I-\sqrt{\Gamma}\left(M-i\omega I\right)^{-1}\sqrt{\Gamma}$)
describe the scattering probabilities of the signal between the cavity
$a_{i}$ and the cavity $a_{j}$ via the mechanical oscillators, and
the noise correlations in the frequency domain have been included.
Thus, the asymmetry of the transmission matrix, i.e. $T_{12}(\omega)\neq T_{21}(\omega)$,
implicates nonreciprocity of the device.

To construct a dissipative coupling path for the two ports, we consider
the MO 2 is strongly damped, with its damping rate $\gamma_{2}$ being
much larger than the decay rates of the cavity modes (assumed to be
equal $\kappa_{1}=\kappa_{2}=\kappa$) and that of the MO 1 (i.e.
$\gamma_{2}\gg\kappa,\gamma_{1}$). We can then adiabatically eliminate
the MO 2 , which leads to dissipative coupling between the two cavity
modes with the strength $Q_{2}=2G_{12}G_{22}/\gamma_{2}$. As a result,
the two cavities are coupled through both a coherent path via MO 1
($a_{1}\rightarrow b_{1}\rightarrow a_{2}$) and a dissipative path
intermediated by MO 2 ($a_{1}\rightarrow b_{2}\rightarrow a_{2}$).
The coupled equations for $a_{1},$ $a_{2}$, and $b_{2}$ become
\begin{equation}
\dot{\mu}^{\prime}(t)=-M^{\prime}\mu^{\prime}(t)+\sqrt{\Gamma^{\prime}}\mu_{\text{in}}^{\prime}(t)-i\sqrt{\Lambda}B_{2,\text{in}},\label{eq:Effective3}
\end{equation}
with $\mu^{\prime}(t)=(a_{1}^{\dagger},a_{2}^{\dagger},b_{1})^{T}$,
$\mu_{\text{in}}^{\prime}(t)=(a_{1,\text{in}}^{\dagger},a_{2,\text{in}}^{\dagger},b_{1,\text{in}})^{T}$,
$\Gamma^{\prime}=diag(\kappa,\kappa,\gamma_{1})$,$\Lambda=diag(\gamma_{1,2},\gamma_{2,2},0)$,
$B_{2,\text{in}}=(b_{2,\text{in}},b_{2,\text{in}},b_{2,\text{in}})^{T}$
and the coefficient matrix

\begin{eqnarray}
M^{\prime} & = & \left[\begin{array}{ccc}
\frac{\kappa-\gamma_{1,2}}{2} & -Q_{2} & -iG_{11}\\
-Q_{2} & \frac{\kappa-\gamma_{2,2}}{2} & -iG_{21}e^{-i\varphi}\\
iG_{11} & iG_{21}e^{i\varphi} & \frac{\gamma_{1}}{2}
\end{array}\right],\label{eq:new matirx}
\end{eqnarray}
where $\gamma_{1,2}=4G_{12}^{2}/\gamma_{2}$ and $\gamma_{2,2}=4G_{22}^{2}/\gamma_{2}$
are corrections of the cavity decay rates induced by the strongly
dissipative MO 2. Using the Fourier transform and the standard input-output
relation again, we get the output vector in the frequency domain 
\begin{equation}
\mu_{\text{out}}^{\prime}(\omega)=S^{\prime}(\omega)\mu_{\text{in}}^{\prime}(\omega)-iL^{\prime}(\omega)B_{2,\text{in}},
\end{equation}
with $S^{\prime}(\omega)=\sqrt{\Gamma^{\prime}}\left(M^{\prime}-i\omega I\right)^{-1}\sqrt{\Gamma^{\prime}}-I$
and $L^{\prime}(\omega)=\sqrt{\Gamma^{\prime}}\left(M^{\prime}-i\omega I\right)^{-1}\sqrt{\Lambda}$.
It follows that the transmission coefficients between the two cavity
modes read
\begin{equation}
S_{12}^{\prime}(\omega)=\frac{\kappa(Q_{1-}+Q_{2})}{D(\omega)},\text{ }S_{21}^{\prime}(\omega)=\frac{\kappa(Q_{1+}+Q_{2})}{D(\omega)},\label{eq:S12_3mode}
\end{equation}
with

\begin{eqnarray}
D(\omega) & = & \left[\frac{\kappa_{1,\text{tot}}}{2}-i(\omega+\omega_{1,1})\right]\left[\frac{\kappa_{2,\text{tot}}}{2}-i(\omega+\omega_{2,1})\right]\nonumber \\
 &  & -\left(Q_{1+}+Q_{2}\right)\left(Q_{1-}+Q_{2}\right),\label{eq:D(w)}
\end{eqnarray}
where we have introduced the frequency $\text{\ensuremath{\omega}}$
dependent corrections induced by the coherent path for the cavity
decay rates $\kappa_{i,\text{tot}}=\kappa-\gamma_{i,1}-\gamma_{i,2}$,
the coupling strengths $Q_{1,\pm}=2G_{11}G_{21}\Sigma(\omega)e^{\pm i\varphi}$
with $\Sigma(\omega)=(\gamma_{1}-i2\omega)^{-1}$, the damping rates
$\gamma_{i,1}=4G_{i1}^{2}|\Sigma(\omega)|^{2}\gamma_{1}$ and the
``optical spring'' frequency shifts $\omega_{i,1}=4G_{i1}^{2}|\Sigma(\omega)|^{2}\omega$.
One should note that the dissipative path by the MO 2 can not induce
such frequency selective properties for the cavity transmission. Consequently,
a non-reciprocal device (with $|Q_{1+}+Q_{2}|\neq|Q_{1-}+Q_{2}|$)
can not be realized at $\omega=0$ by engineering the global phase
$\varphi$, but intriguingly, the signal light may be directionally
amplified or fully isolated for $\omega\neq0$, see the details later.

\begin{figure}
\includegraphics[width=1\columnwidth]{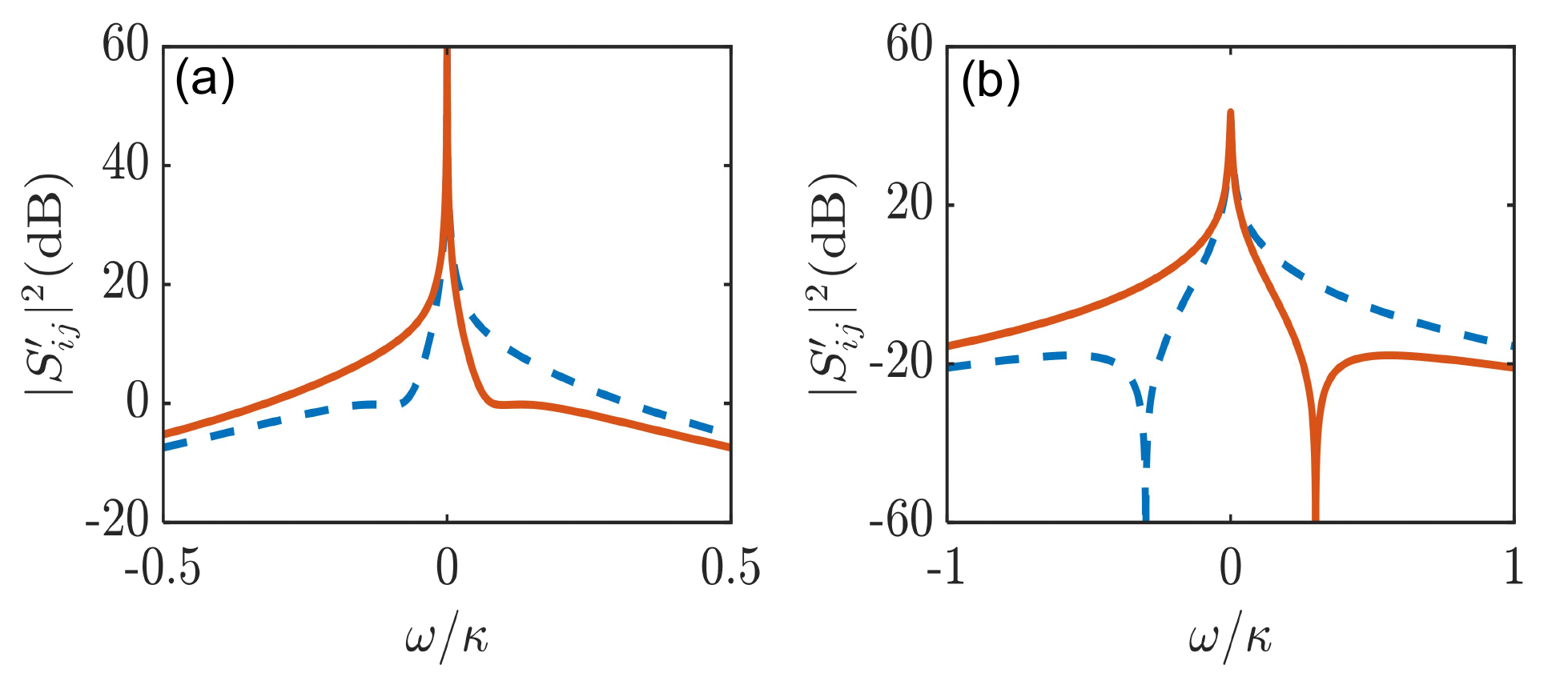}

\caption{\label{fig:S21=000026S12} Scattering probabilities $\left|S_{12}^{\prime}(\omega)\right|^{2}$
(orange solid line) and $\left|S_{21}^{\prime}(\omega)\right|^{2}$
(blue dashed line) as functions of the frequency $\omega$ in units
of $\kappa$ for (a) $(G_{11},G_{12},\gamma_{1},\gamma_{2})/\kappa=(0.13,1.237,0.2,16)$,
$\varphi=-1.25\pi$, and (b) $(G_{11},G_{12},\gamma_{1},\gamma_{2})/\kappa=(0.323,1.198,1,16)$,
$\varphi=0.828\pi$. In both cases we assume $G_{21}=G_{11}$, $G_{22}=G_{12}$,
and $\kappa=1$..}
\end{figure}
%\setboolean{shortarticle}{true}
As the first example in Fig. \ref{fig:S21=000026S12}(a), we show
the scattering probabilities $\left|S_{12}^{\prime}(\omega)\right|^{2}$
and $\left|S_{21}^{\prime}(\omega)\right|^{2}$ in units of dB between
the cavities as a function of $\omega$ for $\varphi=-1.25\pi$. We
look for the regime where the device is configured as a directional
amplifier with lossless nonreciprocal transmission from the opposite
direction, i.e., $\left|S_{21}^{\prime}\right|^{2}>0$ dB and $\left|S_{12}^{\prime}\right|^{2}=0$
dB. Such a regime appears bijective around $\omega/\kappa=\pm0.081$,
where one can achieve $11$ dB of amplification and frequency-insensitive
lossless transmission with respect to the flat local minimum around
$\left|S_{12}^{\prime}\right|^{2}=0$ dB. While the application of
four blue-sideband drive tones enables amplification, the system stability
should be examined carefully and can be checked by using the Routh-Hurwitz
criterion \cite{DeJesus1987}, which imposes the conditions given
by
\begin{eqnarray}
\sum_{j=1,2}4(G_{1j}^{2}+G_{2j}^{2})-\gamma_{1}\gamma_{2}-2\gamma_{2}\kappa-4\kappa^{2} & > & 0,\nonumber \\
\sum_{j=1,2}4(G_{1j}^{2}+G_{2j}^{2})(\gamma_{2/j}+\kappa)-\kappa\gamma_{2}(2\gamma_{1}+\kappa) & > & 0,\nonumber \\
\frac{G_{12}^{2}G_{21}^{2}+G_{11}^{2}G_{22}^{2}}{\gamma_{1}\gamma_{2}\kappa^{2}}-\sum_{j=1,2}\frac{G_{1j}^{2}+G_{2j}^{2}}{4\gamma_{j}\kappa}+\frac{1}{16} & > & 0.
\end{eqnarray}
We have confirmed that the nonreciprocity shown in all figures can
be realized under the stable parameters.
\begin{figure}
\includegraphics[width=1\columnwidth]{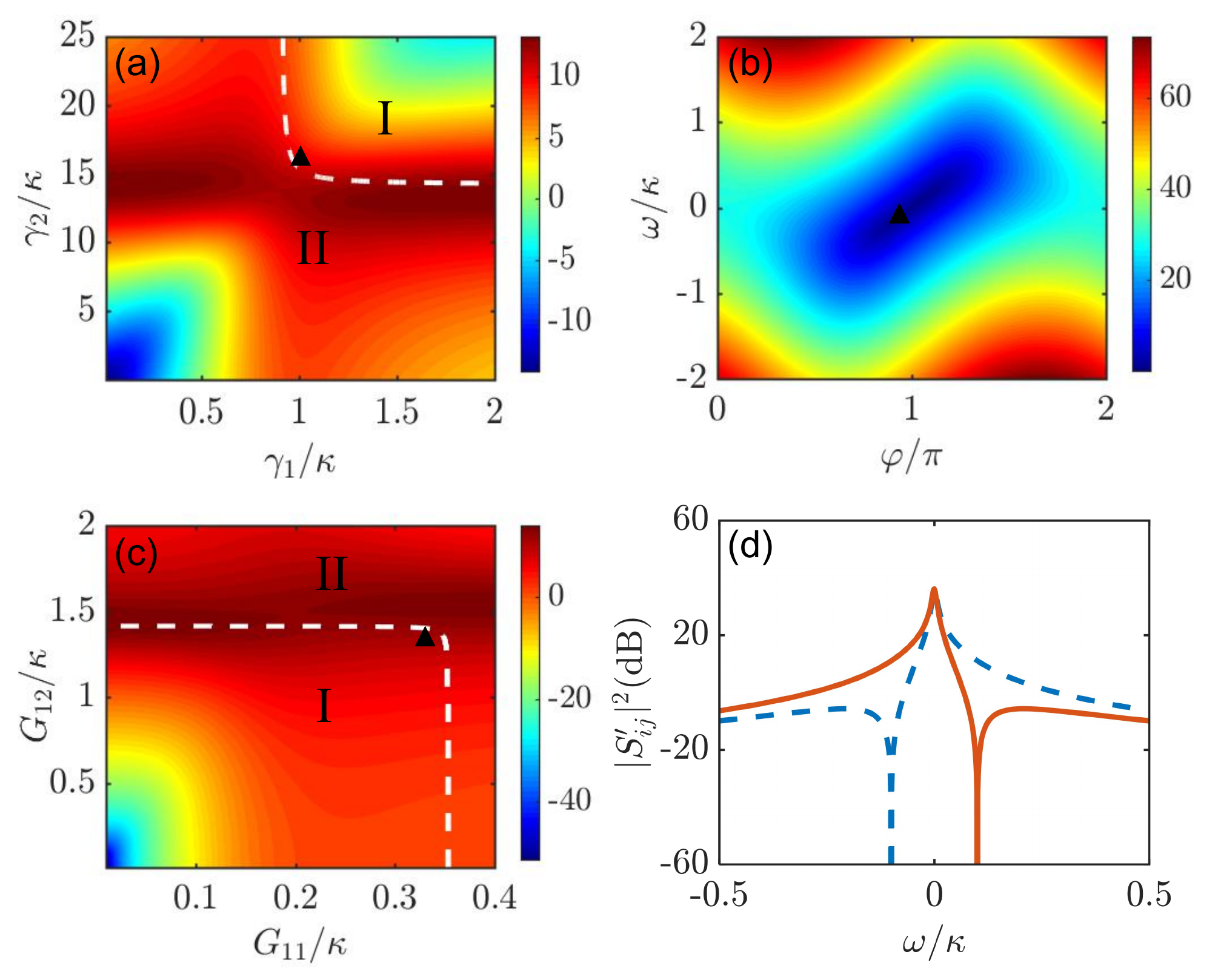}

\caption{\label{fig:Amp=000026ISO} (a) Scattering probability $|S_{21}(\omega)|^{2}$
versus $\gamma_{1}/\kappa$ and $\gamma_{2}/\kappa$ for $(G_{11},G_{12})/\kappa=(0.336,1.333)$,
$\varphi=0.94\pi$, $\omega/\kappa=-0.1$. (b) Numerator of $|S_{12}(\omega)|$
versus the frequency detuning $\omega/\kappa$ and the global phase
$\varphi$ with $(G_{11},G_{12},\gamma_{1},\gamma_{2})/\kappa=(0.336,1.333,1,16).$
(c) Density plot of the scattering probability $\left|S_{21}^{\prime}(\omega)\right|^{2}$
as a function of coupling strengths $G_{11}$ and $G_{12}$. Other
parameters are $(\gamma_{1},\gamma_{2})/\kappa=(1,16)$ and $\varphi\approx0.94\pi$.
(d) Scattering probabilities $\left|S_{12}^{\prime}(\omega)\right|^{2}$
(orange solid line) and $\left|S_{21}^{\prime}(\omega)\right|^{2}$
(blue dashed line) versus $\omega$ for the parameter regime indicated
by the black triangle in (a)-(c). In (a), (c), the white dashed line
indicates the boundary between stable (I) and unstable (II) regions.
In all cases we assume $G_{21}=G_{11},$ $G_{22}=G_{12}$, and $\kappa=1$. }
\end{figure}

Second, the setup with exclusive use of blue-sideband drive tones
can be alternatively configured as an isolator with lossless nonreciprocal
transmission, i.e. $\left|S_{12}^{\prime}\right|^{2}\rightarrow-\infty$
and $\left|S_{21}^{\prime}\right|^{2}=0$ dB in Fig. \ref{fig:S21=000026S12}(b).
Indeed, the absolute isolation from the cavity 2 to the cavity 1 can
be obtained as $Q_{1-}+Q_{2}=0$, giving rise to
\begin{eqnarray}
\frac{\gamma_{1}}{\gamma_{2}} & = & -\frac{G_{11}G_{21}}{G_{12}G_{22}}\text{cos}\varphi,\label{eq:ga1/ga2}
\end{eqnarray}
with $\varphi=\text{atan}(2|\omega|/\gamma_{1})$. Taking the stability
of system into account, we then choose $\gamma_{1}=\gamma_{2}/16=\kappa,$
$G_{11}=G_{21}=0.323\kappa$, $G_{12}=G_{22}=1.198\kappa.$ As such,
the absolute isolation appears at the frequencies $\omega=\gamma_{1}\tan\varphi/2\approx\pm0.3\kappa,$
see Fig. \ref{fig:S21=000026S12}(b), namely the isolation can be
implemented bidirectionally at two working points with the probe detuning
being $\pm0.3\kappa$.

Finally, the nonreciprocal device can be nicely configured as both
an directional amplifier as well as an isolator for the transmission
from the other direction, e.g. $\left|S_{12}^{\prime}\right|^{2}\rightarrow-\infty$
and $\left|S_{21}^{\prime}\right|^{2}\gg0$ dB, and vice versa. To
get the parameter regime (as labeled by the black triangle), we first
show the forward gain $|S_{21}|^{2}$ under the full model Eq. (\ref{eq:S12_3mode})
versus $\gamma_{1}/\kappa$ and $\gamma_{2}/\kappa$ {[}see Fig. \ref{fig:Amp=000026ISO}(a){]},
verifying the effectiveness of the reduced three-mode model (\ref{eq:Effective3})
and the achievable gain under stable conditions. Moreover, we show
that the isolation condition (\ref{eq:ga1/ga2}) is fulfilled at the
phase $\varphi\approx0.94\pi$ and the frequency detuning $\omega=\pm0.1\kappa$
with $(G_{11},G_{12},\gamma_{1},\gamma_{2})/\kappa=(0.336,1.333,1,16)$
{[}see Fig. \ref{fig:Amp=000026ISO}(b){]}. In this regime, $11$
dB of amplification is achieved with the system approaching the boundary
between the stable and unstable region, indicated by the white dashed
line in Fig. \ref{fig:Amp=000026ISO}(a), (c). The frequency-dependence
of the scattering probabilities $\left|S_{12}^{\prime}(\omega)\right|^{2}$
and $\left|S_{21}^{\prime}(\omega)\right|^{2}$ are shown in Fig.
\ref{fig:Amp=000026ISO}(d). 

Furthermore, we calculate the added noise for the amplified output
signal, which arises from the cavity vacuum noise and the mechanical
thermal occupation, and is related to the spectrum density of the
output port (e.g. the cavity $a_{2}$ for $\left|S_{21}^{\prime}\right|^{2}\gg0$)
given by \cite{Malz2018,Jiang2019,Damskagg2019}, 
\begin{align}
S_{\text{out},a_{2}}(\omega)= & \sum_{i=1}^{2}\frac{1}{2}\left|S_{2i}^{\prime}\right|^{2}+\left|S_{23}^{\prime}\right|^{2}(n_{m,1}+\frac{1}{2})\nonumber \\
 & +\sum_{i,j=1}^{2}L_{2i}^{\prime}L_{2j}^{\prime*}(n_{m,2}+\frac{1}{2}),
\end{align}
which is associated with the scattering probabilities $S_{2i}^{\prime}(\omega)$
and $L_{2i}^{\prime}(\omega)$ via the coherent and dissipative paths,
respectively \cite{Caves1982,ClerkRMP2010,Nunnenkamp2014}. The added
noise is then simply given by $\mathcal{N}(\omega)=S_{\text{out},a_{2}}(\omega)/\mathcal{G}-1/2$,
with $\mathcal{G=}\left|S_{21}^{\prime}(\omega)\right|^{2}$. For
the directional amplifier implemented as in Fig. \ref{fig:Amp=000026ISO}(d),
we find that the added noise is $\mathcal{N}(-0.1\kappa)=4.35$ noise
quanta in the absence of thermal mechanical noise, while for the directional
amplifier as in Fig. \ref{fig:S21=000026S12}(a), the added noise
is $\mathcal{N}(\pm0.081\kappa)=8.46$ even for $n_{m,1}=n_{m,2}=3$. 

In conclusion, we have shown that directional amplifier and isolator
can be realized in a single optomechanical setup involving two cavities
and two mechanics  exclusively drive tones on the blue sidebands. The directional
amplification with absolute isolation in the reversal direction appears
at the working points detuned from the resonance, which gives rise
to bi-directional nonreciprocity at two mirrored frequencies for signal
propagating from opposite directions. The amplifier gain can in principle
be enhanced by increasing the strengths of the driving fields, and
is however limited by the stability condition of the system. Such
devices carry great promise for integrated nonreciprocal optical and
microwave devices, as well as the interface between the two frequency
domains.

%\begin{backmatter}
%\bigskip
Acknowledgments - H.W. and S.B.Z. are supported by the National Natural
Science Foundation of China (NSFC) under Grants No. 11774058, No.
11874114, and No. 12174058. Y.L. was supported by the Science Challenge
Project under the Grant No. TZ2018003, and the National NSFC under
Grants No. 11774024, No. 12074030, and No. U1930402.

\smallskip
%\bmsection{Disclosures} The authors declare no conflicts of interest.

%\bigskip

%\end{backmatter}

\bibliography{nonreciprocal1}

%\bibliographyfullrefs{nonreciprocal1}

\end{document}